\documentclass[11pt]{article}
\usepackage{moriond,epsfig}

\def\be{\begin{equation}}
\def\ee{\end{equation}}
\def\bea{\begin{eqnarray}}
\def\eea{\end{eqnarray}}

\def\code{{\tt SUSY\_FLAVOR}}

\newcommand{\gev}{\mbox{GeV}}

\begin{document}
\vspace*{3.5cm}
\title{Rare decays and MSSM phenomenology}

\author{ ANDREAS CRIVELLIN }

\address{Albert Einstein Center for Fundamental Physics -- Institute for Theoretical Physics,\\
University of Bern, CH-3012 Bern, Switzerland}

\maketitle\abstracts{
In this article I review some aspects of flavour phenomenology in the MSSM. 
After an overview of various flavour observables I discuss the constraints on the off-diagonal elements of the squark mass matrices. In this context I present the Fortran code \code~which calculates these processes in the generic MSSM including the complete resummation of all chirally enhanced effects as a new feature of version 2.
Than I discuss where large new physics effects in the MSSM are still possible. As an example of a model which can give large effects in flavour physics I review a model with ``radiative flavour violation'' (RFV) and update the results in the light of the recent LHCb measurement of $B_s\to \mu\mu$. Finally, I recall that the MSSM can generate a sizable right-handed $W$-coupling which affects $B\to\tau\nu$ and can solve the $V_{ub}$ problem.}

\section{Introduction}

In recent years flavour physics has been one of the most active and
fastest developing fields in high energy physics. Numerous new
experiments were carries out but almost all of them reported result in agreement with the Standard Model (SM) predictions (with a few exceptions like the semileptonic CP asymmetry \cite{Lenz:2012az}).

The extensive set of measurements available for rare decays puts
strong constraints on the flavour structure of physics beyond the
Standard Model, in particular, on the 
flavour- and CP- violating parameters of the Minimal Supersymmetric
Standard Model (MSSM), which give rise to additional flavour changing 
neutral currents (FCNCs): they arise from the fact that one cannot (in general) simultaneously
diagonalize the mass matrices of fermions and sfermions. 

The apparent absence of such big effects in flavour (and CP) observables leads to the conclusion that the MSSM
couplings which generate FCNCs (and CP) violation must actually be
strongly suppressed \cite{Bertolini:1990if}.  The difficulty to explain this suppression is
known as the ``SUSY flavour problem'' and the ``SUSY CP problem''.
\footnote{Even if one adopts the so-called Minimal Flavour Violation (MFV)
hypothesis \cite{D'Ambrosio:2002ex}, which requires that {\it all} FCNC effects
originate from the Yukawa couplings of the superpotential,
supersymmetric contributions to various flavour and CP-violating
amplitudes can still be of comparable (or sometimes even much larger,
like in the case of the electron and neutron EDMs or
$B_s\to\mu^+\mu^-$) size as the corresponding SM contributions.}

For studying the various constraints from flavour observables the Fortran code \code~\cite{susyflavour} was developed. It is a universal computational tool which can calculate the flavour observables listed in Table~1 in the generic MSSM . One important new feature of \code~v2 is that it includes the resummation of all chirally enhanced corrections (including all effects from flavour non-diagonal terms) using the results of Ref.~\cite{Crivellin}. This extends to applicability of \code~to regions in parameter space with large values of $\tan\beta$ and/or large trilinear SUSY breaking terms.

Table~1 also gives an overview which off-diagonal elements of the sfermion mass matrices $\Delta^{q\,AB}_{ij}=\delta^{q\,AB}_{ij}\sqrt{m_{\tilde q^A_i}^2m_{\tilde q^B_j}^2}$ are typically most stringently constrained by which process. In the down-squark sector, the constraints on $\Delta^{d\,AB}_{ij}$ range from $\rm{Im}[\Delta^{d\,LR}_{12}]< {\cal{O}}( 10^{-4})$ to $\Delta^{d\,RR}_{23}<{\cal{O}}( 10^{-1})$ and in the lepton sector from $\Delta^{\ell\,LR}_{12}< {\cal{O}}( 10^{-5})$ to $\Delta^{\ell\,RR}_{23}< 1$  for typical SUSY masses (see for example Ref.~\cite{Altmannshofer:2009ne} for a recent overview). In the up sector, only the elements $\delta^{u\;AB}_{12}$ are severely constrained from D mixing and since the LL-elements are connected via the SU(2) relation, the constraints from the down-sector transfer to the up sector (an exception is the case when the squark mass matrix is exactly aliened to $Y^d Y^{d\dagger}$ \cite{Crivellin:2010ys}).

\begin{table}[htbp]
\renewcommand{\arraystretch}{1.45}
\begin{center}
\begin{tabular}{|lcr|}
\hline
Observable & Most stringent constraints on &Experiment \\ \hline\hline
\multicolumn{3}{|l|}{$\Delta F=0$} \\ \hline\hline

$\frac{1}{2}(g-2)_e$ & $\rm{Re}\left[\delta^{\ell\,LR,RL}_{11}\right]$  &$(1 159 652 188.4 \pm4.3) \times 10^{-12}$ \\

$\frac{1}{2}(g-2)_\mu$ & $\rm{Re}\left[\delta^{\ell\,LR,RL}_{22}\right]$ &$(11659208.7\pm8.7)\times10^{-10}$ \\

$\frac{1}{2}(g-2)_\tau$ & $\rm{Re}\left[\delta^{\ell\,LR,RL}_{33}\right]$ &$<1.1\times 10^{-3}$ \\

$|d_{e}|$(ecm) & $\rm{Im}\left[\delta^{\ell\,LR,RL}_{11}\right]$ &$<1.6 \times 10^{-27}$ \\

$|d_{\mu}|$(ecm) & $\rm{Im}\left[\delta^{\ell\,LR,RL}_{22}\right]$ &$<2.8\times 10^{-19}$ \\

$|d_{\tau}|$(ecm) & $\rm{Im}\left[\delta^{\ell\,LR,RL}_{33}\right]$ &$<1.1\times 10^{-17}$ \\

$|d_{n}|$(ecm) & $\rm{Im}\left[\delta^{d\,LR,RL}_{11}\right]$, $\rm{Im}\left[\delta^{u\,LR,RL}_{11}\right]$ &$<2.9 \times 10^{-26}$\\ \hline\hline

\multicolumn{3}{|l|}{$\Delta F=1$}\\ \hline \hline

$\mathrm{Br}(\mu\to e \gamma)$ & $\delta^{\ell\,LR,RL}_{12,21}$, $\delta^{\ell\,LL,RR}_{12}$ & $<2.8 \times 10^{-11}$\\ 

$\mathrm{Br}(\tau\to e \gamma)$ & $\delta^{\ell\,LR,RL}_{13,31}$, $\delta^{\ell\,LL,RR}_{13}$ & $<3.3\times 10^{-8}$\\ 

$\mathrm{Br}(\tau\to \mu \gamma)$ & $\delta^{\ell\,LR,RL}_{23,32}$, $\delta^{\ell\,LL,RR}_{23}$ & $<4.4\times 10^{-8}$\\ 

$\mathrm{Br}(K_{L }\to \pi^{0} \nu \nu)$ & $\delta^{u\,LR}_{23}, \delta^{u\,LR}_{13}\times\delta^{u\,LR}_{23}$ & $< 6.7\times10^{-8}$ \\

$\mathrm{Br}(K^{+}\to \pi^{+} \nu \nu)$ & sensitive to $\delta^{u\,LR}_{13}\times\delta^{u\,LR}_{23}$ & $17.3^{+11.5}_{-10.5}\times 10^{-11}$ \\

$\mathrm{Br}(B_{d}\to e e)$ & $\delta^{d\,LL,RR}_{13}$ & $<1.13\times 10^{-7}$\\

$\mathrm{Br}(B_{d}\to \mu \mu)$ & $\delta^{d\,LL,RR}_{13}$ & $<1.8\times 10^{-8}$\\

$\mathrm{Br}(B_{d}\to \tau \tau)$ & $\delta^{d\,LL,RR}_{13}$ & $<4.1\times10^{-3}$ \\

$\mathrm{Br}(B_{s}\to e e)$ & $\delta^{d\,LL,RR}_{23}$ & $<7.0\times 10^{-5}$\\

$\mathrm{Br}(B_{s}\to \mu \mu)$ & $\delta^{d\,LL,RR}_{23}$ & $<1.08\times 10^{-8}$\\

$\mathrm{Br}(B_{s}\to \tau \tau)$ & $\delta^{d\,LL,RR}_{23}$ & $--$\\

$\mathrm{Br}(B_{s}\to \mu e)$ & $\delta^{d\,LL,RR}_{23}\times\delta^{\ell\,LL,RR}_{12}$ & $<2.0\times 10^{-7}$\\

$\mathrm{Br}(B_{s}\to \tau e )$ & $\delta^{d\,LL,RR}_{23}\times\delta^{\ell\,LL,RR}_{13}$ & $<2.8\times 10^{-5}$\\

$\mathrm{Br}(B_{s}\to \mu \tau)$ & $\delta^{d\,LL,RR}_{23}\times\delta^{\ell\,LL,RR}_{23}$ & $<2.2\times 10^{-5}$\\

$\mathrm{Br}(B^+\to \tau^+ \nu)$ & -- & $(1.65\pm 0.34)\times10^{-4}$ \\

$\mathrm{Br}(B_{d}\to D\tau \nu)/\mathrm{Br}(B_{d}\to Dl \nu)$ & -- &
($0.407 \pm 0.12 \pm 0.049)$ \\

$\mathrm{Br}(B\to X_{s} \gamma)$ & $\delta^{d\,LL,RR}_{23}$ for large $\tan\beta$, $\delta^{d\,LR}_{23,32}$   & $(3.52\pm 0.25) \times
10^{-4}$\\

\hline\hline
\multicolumn{3}{|l|}{$\Delta F=2$}\\ \hline \hline
$|\epsilon_{K}|$ & $\rm{Im}\left[(\delta^{d\,LL,RR}_{12})^2\right]$, $\rm{Im}\left[(\delta^{d\,LR}_{12,21})^2\right]$ & $(2.229 \pm 0.010)\times 10^{-3}$ \\

$\Delta M_{K}$ & $\delta^{d\,LL,RR}_{12}$, $\delta^{d\,LR}_{12,21}$ & $(5.292 \pm 0.009)\times10^{-3}~\mathrm{ps}^{-1}$\\

$\Delta M_{D}$ & $\delta^{u\,LL,RR}_{12}$, $\delta^{u\,LR}_{12,21}$ & $(2.37^{+0.66}_{-0.71}) \times10^{-2}~\mathrm{ps}^{-1}$\\

$\Delta M_{B_{d}}$ & $\delta^{d\,LL,RR}_{13}$, $\delta^{d\,LR}_{13,31}$ & $(0.507 \pm0.005)~\mathrm{ps}^{-1}$\\

$\Delta M_{B_{s}}$ & $\delta^{d\,LL,RR}_{23}$, $\delta^{d\,LR}_{23,32}$ & $(17.77 \pm0.12)~\mathrm{ps}^{-1}$\\ \hline
\end{tabular}

\end{center}
\caption{List of observables calculated by \code{} v2 and their currently measured values or 95\% CL (except for $\mathrm{Br}(B_{d}\to e^+ e^-)$ and $\mathrm{Br}(B_{d}\to \tau^+ \tau^-)$ for which the 90\% C.L bounds are given). We also give the off-diagonal elements of the sfermion mass matrices which are most stringently constrained by the corresponding process. \label{tab:proc}}

\end{table}

\section{Where are deviations from the SM still possible?}

From table~1 we see that all off-diagonal elements of the down-squark mass matrix are constrained while in the up-sector $\delta^{u\;RR,LR,RL}_{13,23}$ can be large \footnote{Recently it has been pointed out that $B\to K^{(\star)} \ell^+\ell^-$ constrains $\delta^{u\;LR}_{23}< {\cal{O}}( 10^{-1})$ \cite{Behring:2012mv}}. However, the effect of $\delta^{u\;RR}_{13,23}$ on flavour-observables is very limited and thus we focus on $\delta^{u\;LR,RL}_{13,23}$. While sizable values for $\delta^{u\;LR}_{13,23}$ are needed in models with radiative flavour violation (RVF) if the CKM matrix is generated in the up-sector, $\delta^{u\;LR}_{31}$ can generate a sizable right-handed $W$ coupling which affects $B\to \tau \nu$ and $B\to \pi \ell \nu$.

\subsection{Radiative flavour violation}

An interesting alternative to MFV, which can still give interesting effects in flavour observables, is the MSSM with radiative flavour violation (RFV) \cite{RFV,RFV2}. RFV means that the CKM matrix is the unit matrix at tree-level and all off-diagonal elements arise from SUSY loop diagrams. 

\begin{figure}
\includegraphics[width=0.44\textwidth]{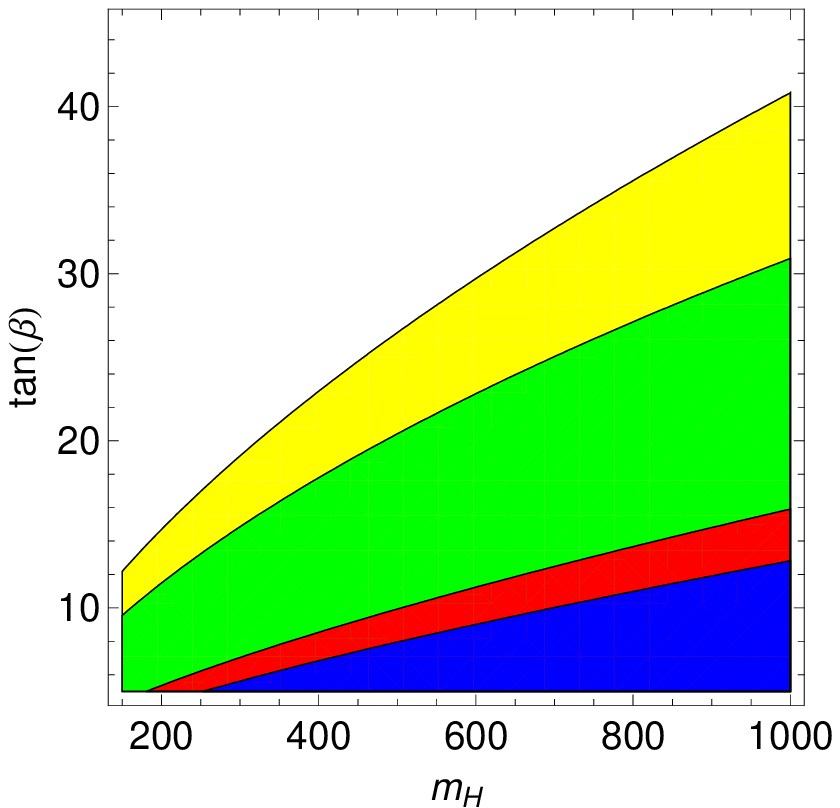}
\includegraphics[width=0.46\textwidth]{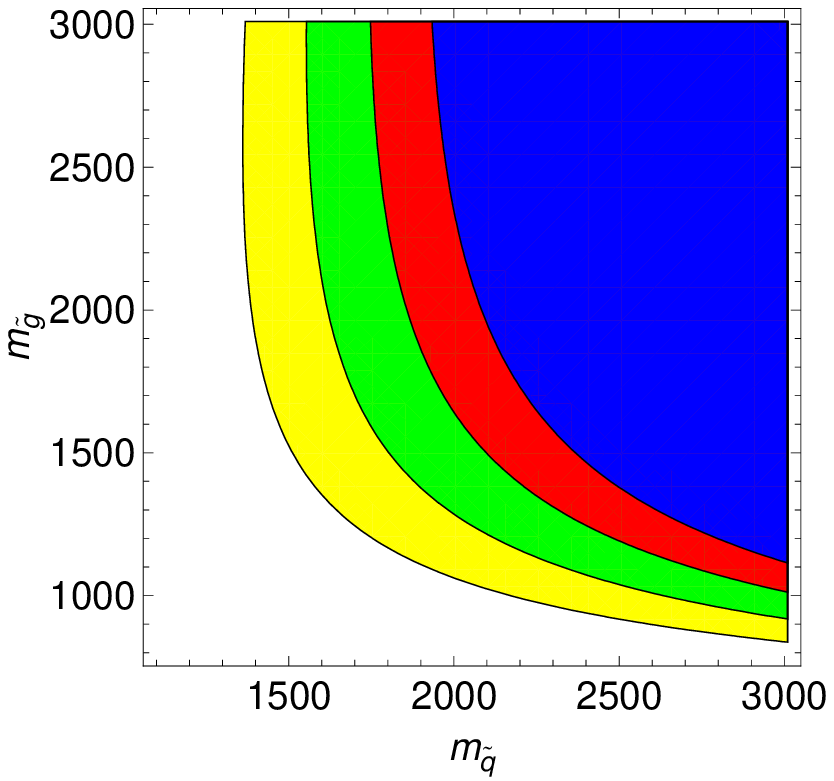}
\caption{
Left: Allowed region in the $m_{H}$--$\tan\beta$ plane for different
values of $\epsilon_b$ from {$\rm{Br}[B_s\to\mu^+\mu^-]\leq 4.5\cdot
10^{-9}[95\% CL] $}. 
Yellow: $\epsilon_b=0.005$, green: $\epsilon_b=0.01$, 
red: $\epsilon_b=-0.005$, blue: $\epsilon_b=-0.01$ (light to dark). Note that also destructive interference with the SM can occur, depending on the sign of $\mu$.
Right: Allowed regions in the $m_{\tilde q}-m_{\tilde g}$ plane assuming that the CKM matrix is generated in the up-sector. Constraints from Kaon mixing for different values of $M_2$ assuming that the CKM matrix is generated in the up sector. Yellow(lightest): $M_2=1000 \gev$, green: $M_2=750\gev$, red: $M_2=500\gev$ and blue(darkest): $M_2=250\gev$.
\label{Bmumu_Kaon}}
\end{figure}

If the CKM matrix is generated in the down sector, constraints on the SUSY masses from $b\to s \gamma$ arise. In addition, $B_s\to \mu^+\mu^-$ can be enhanced or suppressed compared to the SM prediction depending on the sign of $\mu$ (see Ref.~\cite{RFV2} for details). We take the opportunity to update the left plot in Fig.~\ref{Bmumu_Kaon} using the new LHCb result \cite{LHCb} and find that still a large region parameter space is compatible with the new stringent constraints from $B_s\to \mu^+\mu^-$. 

\begin{figure}
\centering
\includegraphics[width=0.6\textwidth]{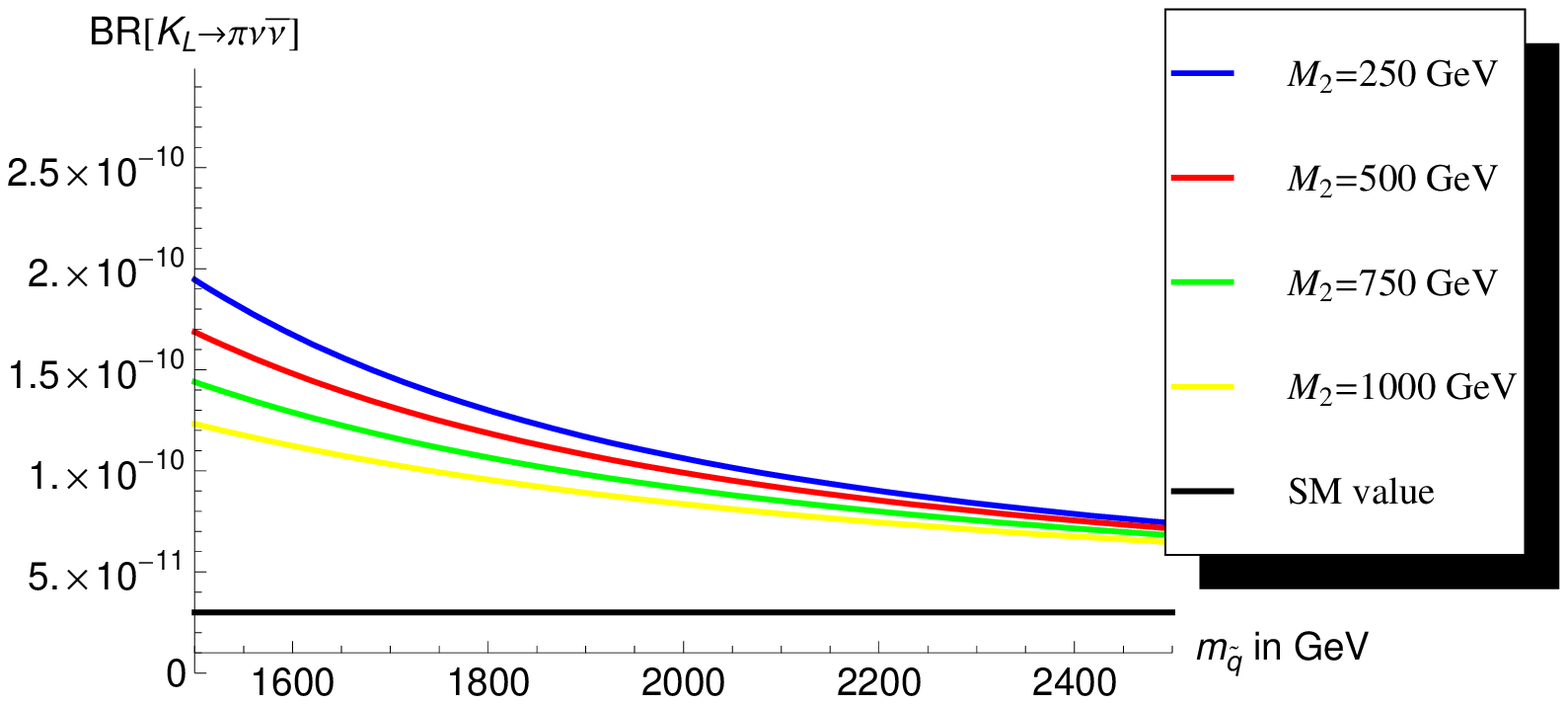}
\includegraphics[width=0.6\textwidth]{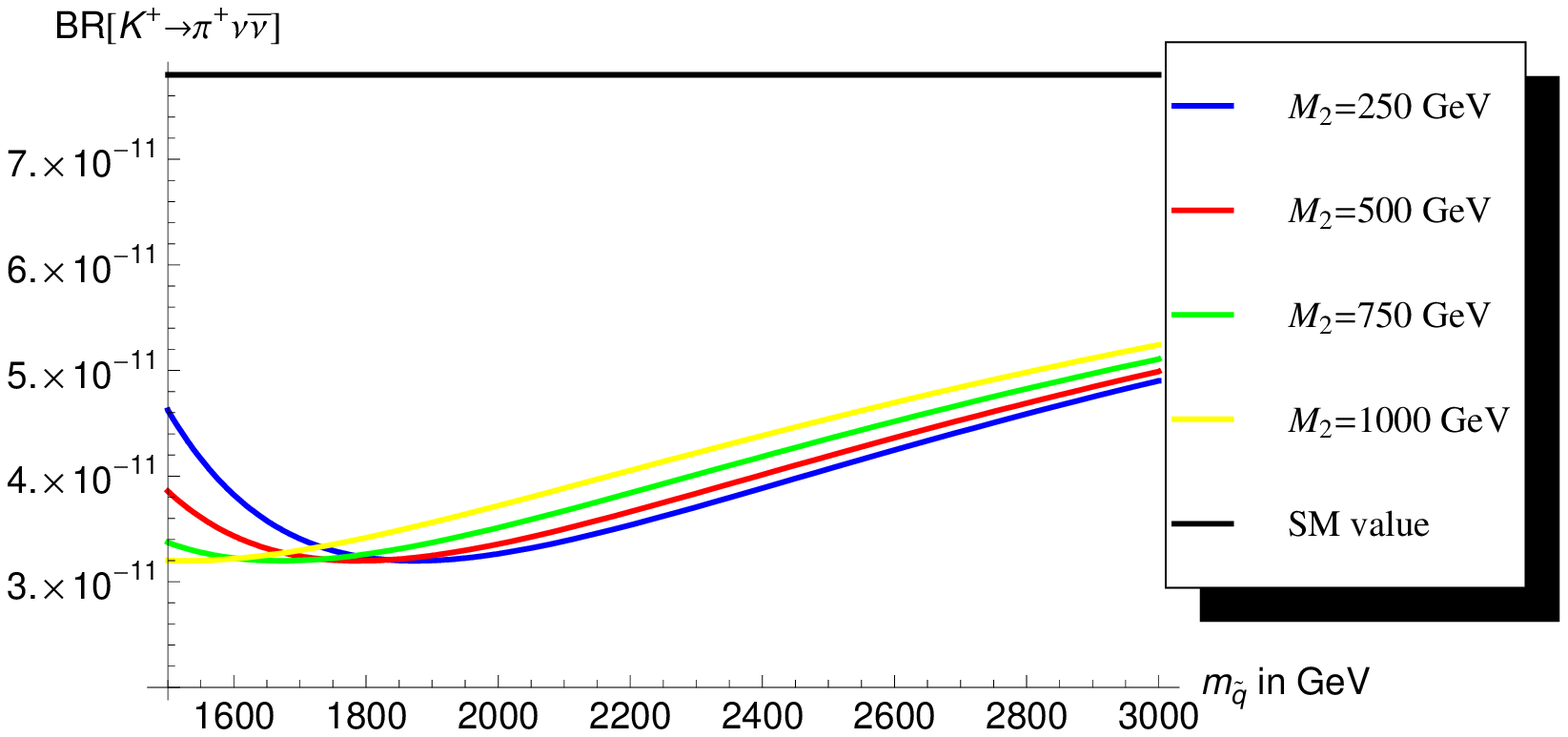}
\caption{Predicted branching ratio for the rare Kaon decay $K_L\to\pi\nu\overline{\nu}$ (upper plot) and $K^+\to\pi^+\nu\overline{\nu}$ (lower plot) assuming that the CKM matrix is generated in the up-sector for $m_{\tilde{q}}=m_{\tilde{g}}$. The branching ratio is enhanced due to a constructive interference with the SM contribution.
  \label{Klongtopinunu}}
\end{figure}

If the CKM matrix is generated in the up-sector, sizable values for $\delta^{u\;LR}_{13,23}$ are needed. In this case constraints from Kaon mixing (and $B\to K^{(\star)} \ell^+\ell^-$ \cite{Behring:2012mv}) arise (see right plot in Fig.~\ref{Bmumu_Kaon}). In addition, $K_L\to \pi\nu\nu$ and $K^+\to \pi^+\nu\nu$ receives sizable contributions from chargino-$Z$ penguins. We see from Fig.~\ref{Klongtopinunu} that RFV with CKM generation in the up-sector predicts an enhancement (suppression) of $K_L\to \pi\nu\nu$ ($K^+\to \pi^+\nu\nu$) with respect to the standard model prediction.

\subsection{Right-handed W coupling}

\begin{figure}
\includegraphics[width=0.62\textwidth]{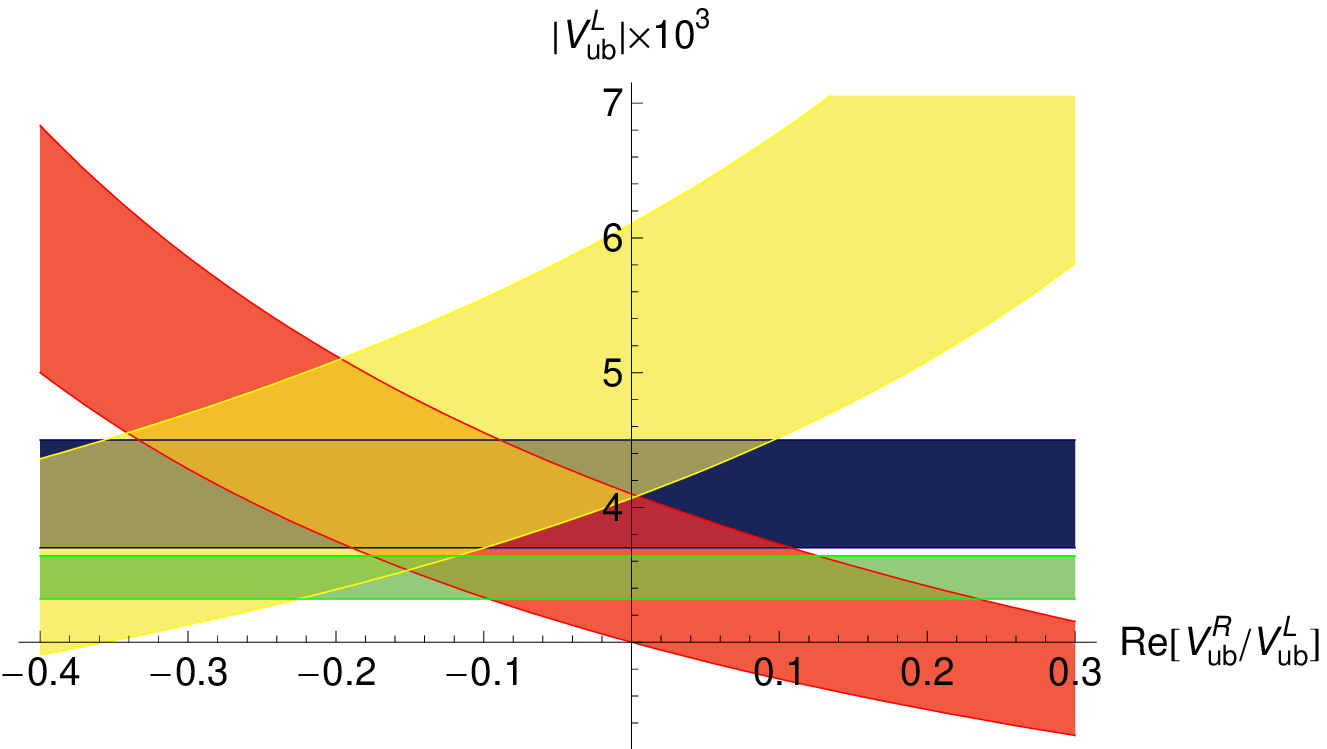}
\includegraphics[width=0.36\textwidth]{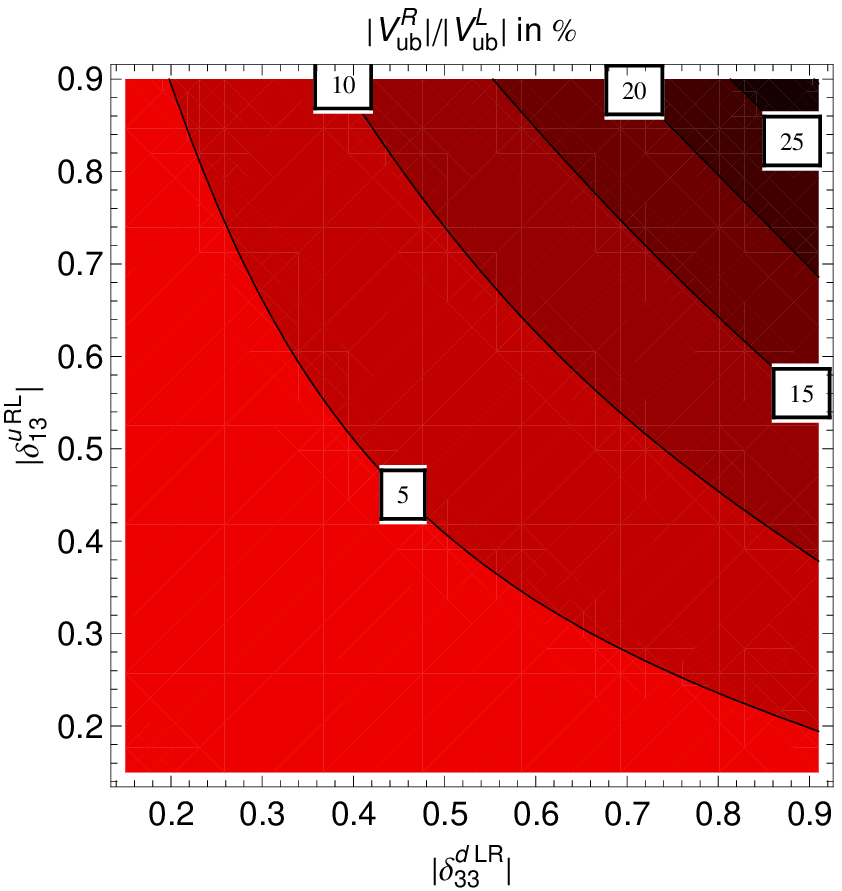}
\vspace{5mm}
\caption{Left: $\left|V^L_{ub}\right|$ as a function of
  $\rm{Re}\left[V^R_{ub}/V^L_{ub}\right]$ extracted from different
  processes. Blue(darkest): inclusive decays. Red(gray): $B\to \pi l\nu$.
  Yellow(lightest gray): $B\to \tau\nu$. Green(light gray): $V^L_{ub}$ determined from CKM unitarity.
Right: Relative strength of the induced right-handed
  coupling $|V^R_{ub}|$ with respect to $|V^L_{ub}|$
  for $M_{\rm{SUSY}}=1\,\rm{TeV}$. $|V^L_{ub}|$ is determined from CKM unitarity.}
  \label{RCKM}
\end{figure}

$\delta^{u\;LR}_{31}$ in combination with $\delta^{d\;LR}_{33}$ can induce a sizable right-handed $W$-coupling to up and bottom~\cite{Crivellin:2009sd}. As we see from the right plot of Fig.~\ref{RCKM}, the strength of the right-handed coupling can reach about 10\% of the left-handed one. Such a large right-handed admixture $V_{ub}^R$ changes the Feynman rule for the $W$-up-bottom vertex to $\frac{ - i g_2 \gamma^\mu}{\sqrt 2 }
 \left( {V_{fi}^L P_L  + V_{fi}^R P_R } \right)$
and alters the determination of the left-handed SM coupling $V_{ub}^L$ from in and exclusive leptonic and semileptonic $B$ decays. The left plot of Fig.~\ref{RCKM} shows that for $V_{ub}^R=0$ there is a 2.8 $\sigma$ discrepancy between the value of $V_{ub}^L$ obtained a fit using CKM unitarity and the one extracted from $B\to\tau\nu$ (see Moriond update of Ref.~\cite{Hocker:2001xe}). This discrepancy can be removed by a small admixture of $V_{ub}^R$ with opposite sign compared to $V_{ub}^R$.

\section{Conclusions}

In these proceedings I briefly reviewed flavour phenomenology in the generic MSSM. Many flavour observables put stringent constraints on the off-diagonal elements of the squark mass matrices. For the calculation of theses constraints \code{} is a useful tool and since v2 now contains the complete resummation of all chirally enhanced effect it can be used also for regions in parameter space with large values of $\tan\beta$ and/or large trilinear $A$-terms.  

While all flavour off-diagonal elements of the down-squark mass matrix are constrained, the bounds on the elements of the up-squark mass matrix involving the third generation are much less stringent. This allows for sizable effects in $K_L\to\pi\nu\nu$ and $K^+\to\pi^+\nu\nu$ (for example in models with RFV if the CKM matrix is generated in the up sector) and using $\delta^{u\;LR}_{31}$ a right-handed $W$-coupling to up and bottom can be induced via loops. Such a right-handed $W$-coupling can enhance $B\to\tau\nu$ with respect to the SM prediction and bring the determination of $V_{ub}$ from CKM unitarity and from $B\to\tau\nu$ into agreement.

\section*{Acknowledgments}

I thank the organizers, especially Stefan Pokorski, for the invitation and the possibility to present these results. This work is supported by the Swiss National Science Foundation. The Albert Einstein Center for Fundamental Physics is supported by the ``Innovations- und Kooperationsprojekt C-13 of the Schweizerische Universit\"atskonferenz SUK/CRUS''. I thank Ulrich Nierste for proofreading the article.

\end{document}